\documentclass[3p,twocolumn,times]{elsarticle}

\usepackage{amsmath}
\usepackage{amsfonts}
\usepackage{multirow}
\usepackage{hyperref}
\usepackage{epsfig}
\newcommand{\be}{\begin{equation}}
\newcommand{\ee}{\end{equation}}
\newcommand{\bea}{\begin{eqnarray}}
\newcommand{\eea}{\end{eqnarray}}
\newcommand{\rbox}{\rule[-0.25cm]{0cm}{8mm}}

\allowdisplaybreaks

\begin{document}

\begin{frontmatter}

\title{A new method to study the number of colors in the final-state interactions of hadrons }

\author{Ling-Yun Dai$^{1}$}
\ead{l.dai@fz-juelich.de}
\author{Ulf-G. Mei{\ss}ner$^{2,1}$}
\ead{meissner@hiskp.uni-bonn.de}
\address{$^1$ Institut f\"ur Kernphysik, Institute for Advanced Simulation and
J\"ulich Center for Hadron Physics, Forschungszentrum J{\"u}lich, D-52425 J{\"u}lich, Germany}
\address{$^2$ Helmholtz Institut f\"ur Strahlen- und Kernphysik and Bethe Center
 for Theoretical Physics, Universit\"at Bonn, D-53115 Bonn, Germany}

\begin{abstract}
Based on Chiral Perturbation Theory we introduce the dependence on the number
of colors ($N_C$) for the $\pi\pi\to\pi\pi$ scattering amplitudes.
Those amplitudes are calculated from dispersion relations that respect analyticity
and coupled channel unitarity,  as well as
accurately describing experiment. By varying $N_C$ the trajectories of the poles and
residues (the couplings to
$\pi\pi$) of the light mesons, the $\sigma/f_0(500)$, $f_0(980)$, $\rho(770)$ and
$f_2(1270)$ are investigated. Our results show that the method proposed is a
reliable way to study the $N_C$ dependence in hadron-hadron scattering
with final-state interactions.
\end{abstract}



\end{frontmatter}

The lightest scalar mesons are rather interesting as they have the same quantum numbers as
the QCD vacuum. The nature of them is still a mystery~\cite{MRP07Rev}-\cite{Pelaez15}.
The phenomenology of these is  complicated due the contribution from important
final-state interactions (FSI)~\cite{AMP-FSI,Meissner:1990kz}.
Dispersion relations are the natural way to include FSI, see e.g.~\cite{Kang:2013jaa,Chen2015}.
For some of the light mesons, like the $\sigma$, $\kappa$, their existence has been
confirmed~\cite{zheng00}-\cite{Moussallam06} 
and accurate pole locations and $\pi\pi$ couplings, including also the $\rho(770)$,
have been given in
Refs.~\cite{caprini06,Descotes04,PelaezPRL}.
Concerning the nature of the scalar mesons, there are a cornucopia of models
\cite{Jaffe:1976ig}-\cite{Wilson2016}.
Among them tracking the large $N_C$ trajectories of the poles is an effective diagnostics
to distinguish ordinary from  non-ordinary
quark-antiquark structure as considered
in~\cite{Pelaez04}-\cite{DLY11}.
However, these analyses based on unitarized Chiral Perturbation theory (U$\chi$PT)
lack crossing symmetry. Unitarization itself
will also generate spurious poles~\cite{Qin:2002hk} and cuts \cite{Pelaez02,DLY11}.
In contrast, dispersion relations respect analyticity, but including coupled channel unitarity
and the $N_C$ dependence is difficult.
Clearly, both analyticity and coupled channel unitarity are critical in the region of the
$\overline{K}K$ threshold, close to which the $f_0(980)$ is located. To solve this problem, we
use an Omn\`es representation based on the phase of the relevant amplitudes,
rather than the elastic phase shift~\cite{DLY-MRP14,DLY2015}.
There has been renewed interest in the study of the  large $N_C$
limit~\cite{'tHooft:1973jz,'tHooft:1974hx}
of the properties of resonances~\cite{Witten:1979kh,Coleman1985,Cohen1998}.
Weinberg~\cite{Weinberg2013} pointed out that resonant tetraquark states could exist due to the
contribution of the leading order (LO) \lq connected' diagrams to the Green functions.
Their widths are $\mathcal{O}(N_C^{-1})$, as narrow as ordinary mesons.
They could be even narrower, with width of $\mathcal{O}(N_C^{-2})$, when the flavor
of the quarks is combined in different ways~\cite{Knecht2013}. The width could
also be wide as $\mathcal{O}(1)$, see  \cite{Cohen:2014vta}.
There are many other interesting discussions such as~\cite{Lebed2013,Cohen2014}
and references therein.
In this paper we focus on establishing a \lq practical' way to study the $N_C$ dependence of
the scattering amplitudes, built into dispersion relations.
Resonances appearing in the intermediate states are also studied.

In this letter we first use dispersive methods to obtain the $\pi\pi$ scattering amplitudes
up to 2~GeV. We construct the amplitudes in a model-independent way,   which is both analytic
and respects coupled channel unitarity.
We also recalculate the analytical expressions of  the $IJ=00,02,11$ waves in $SU(3)$
Chiral Perturbation Theory ($\chi$PT) up to $\mathcal {O}(p^4)$.
By matching with the $\chi$PT amplitudes up to $\mathcal{O}(N_C^{-1})$,
we introduce the $N_C$-dependence into the dispersive amplitudes. This $N_C$ dependence is
automatically transferred to the
high-energy region, where the FSI are implemented by a dispersion relation.
We give the trajectories of the poles and residues by varying $N_C$.
The behavior of the $\rho(770)$, $f_2(1270)$, $\sigma/f_0(500)$ and $f_0(980)$ show that this
is a reliable way to study the number of colors in hadron-hadron scattering.
The $N_C$ trajectory of
the light scalar mesons supports a mixed structure of hadronic molecule and
$\bar{q}q$ components (for a recent review on hadronic molecules, see Ref.~\cite{Guo:2017jvc}),
while a tetra-quark component is also possible.

We first present our $IJ=00,02,11$ partial waves of $\pi\pi\to\pi\pi$ calculated
in a model-independent way. We start from:
\be\label{eq:TP}
T^I_J(s)=P^I_J(s)\Omega^I_J(s).
\ee
where $\Omega^I_{J}(s)$ is the Omn\`es function~\cite{Omnes1958}:
\be\label{eq:Omnes}
\Omega^I_{J}(s)=\exp\left(\frac{s}{\pi} \int^\infty_{s_{th}} ds' \frac{\varphi^I_{J}(s')}{s'(s'-s)}\right) \,\, .
\ee
with $\varphi^I_{J}(s)$ the phase of the partial wave amplitude $T^I_{J}(s)$, which has been given in previous amplitude analysis~\cite{DLY-MRP14,DLY2015}.
This phase is known from experiment up to roughly 2~GeV, while in the higher energy
region it is constrained by unitarity. The Omn\`es function is truncated at
$s=22$~GeV$^2$. The function $P^I_J(s)$ includes
the effect of the left-hand cut (l.h.c) and corrections that come from the distant right-hand
cut (r.h.c) above 2~GeV. The latter one has a tiny contribution to the region where  the light
mesons appear.
Other information is provided by chiral dynamics that fixes the Adler zero in the S-wave,
and the approach to the threshold of the S-, P-, and D-waves in terms of scattering lengths and
effective ranges. We therefore parameterize the $P^I_J(s)$ as
\be\label{eq:P}
P^I_J(s)=(s-z^I_J)^{n_J}\sum_{k=1}^n {\alpha^I_J}_k (s-4M_\pi^2)^{k-1}\,,
\ee
where $z^I_J$ is the Adler zero for the S-wave and $4M_\pi^2$ for P- and  D-waves.
The parameter $n_J$ is 1 for S- and P-waves and 2 for D-waves.
The parameters $\alpha_i$ are given in Table~\ref{tab:para;alpha}. The units of the
$\alpha_k$ are chosen to ensure that the amplitude $T^I_J(s)$ is dimensionless.
\begin{table}[htbp]
\hspace{-1.5cm}
\vspace{-0.0cm}
{\footnotesize
\begin{center}
\tabcolsep=0.11cm
\begin{tabular}  {|c|c|c|c|c|c|c|}
\hline
                                                                & ${T_{11}}^0_S$  &  ${T_{11}}^0_D$ & ${T_{11}}^1_P$ \rbox \\[0.5mm] \hline
\rule[0.32cm]{0.1cm}{0cm}$\alpha_1$\rule[0.32cm]{0.1cm}{0cm}    & 2.4051          & 0.2972          & 0.4283           \\ \hline
\rule[0.32cm]{0.1cm}{0cm}$\alpha_2$\rule[0.32cm]{0.1cm}{0cm}    & $-$1.9451      & $-$0.9354         &-0.2976           \\ \hline
\rule[0.32cm]{0.1cm}{0cm}$\alpha_3$\rule[0.32cm]{0.1cm}{0cm}    & 1.3473(36)      &  2.1931(2)      & 0.6173(16)       \\ \hline
\rule[0.32cm]{0.1cm}{0cm}$\alpha_4$\rule[0.32cm]{0.1cm}{0cm}    & $-$0.4629(19)   & $-$2.6108(2)      &$-$0.7092(11)       \\ \hline
\rule[0.32cm]{0.1cm}{0cm}$\alpha_5$\rule[0.32cm]{0.1cm}{0cm}    & 0.0038(6)       &  1.6508(1)      & 0.3774(4)        \\ \hline
\rule[0.32cm]{0.1cm}{0cm}$\alpha_6$\rule[0.32cm]{0.1cm}{0cm}    & 0.0307(2)     & $-$0.5679(1)      & $-$0.0909(1)       \\ \hline
\rule[0.32cm]{0.1cm}{0cm}$\alpha_7$\rule[0.32cm]{0.1cm}{0cm}    &$-$0.0045(1)     &  0.1004(1)      & 0.0081(1)        \\ \hline
\rule[0.32cm]{0.1cm}{0cm}$\alpha_8$\rule[0.32cm]{0.1cm}{0cm}    & $-$            & $-$0.0071(1)      & $-$                \\ \hline
\end{tabular}
\caption{\label{tab:para;alpha}The fit parameters  as given in Eq.~(\ref{eq:P}).
The errors are given by MINUIT and notice that the $\alpha_{1,2}$ are
fixed by the scattering lengths and slope parameters~\cite{DLY2015,KPY,Ke4}.}
\end{center}
}
\end{table}

\begin{figure}[th]
\includegraphics[width=0.48\textwidth,height=0.45\textheight]{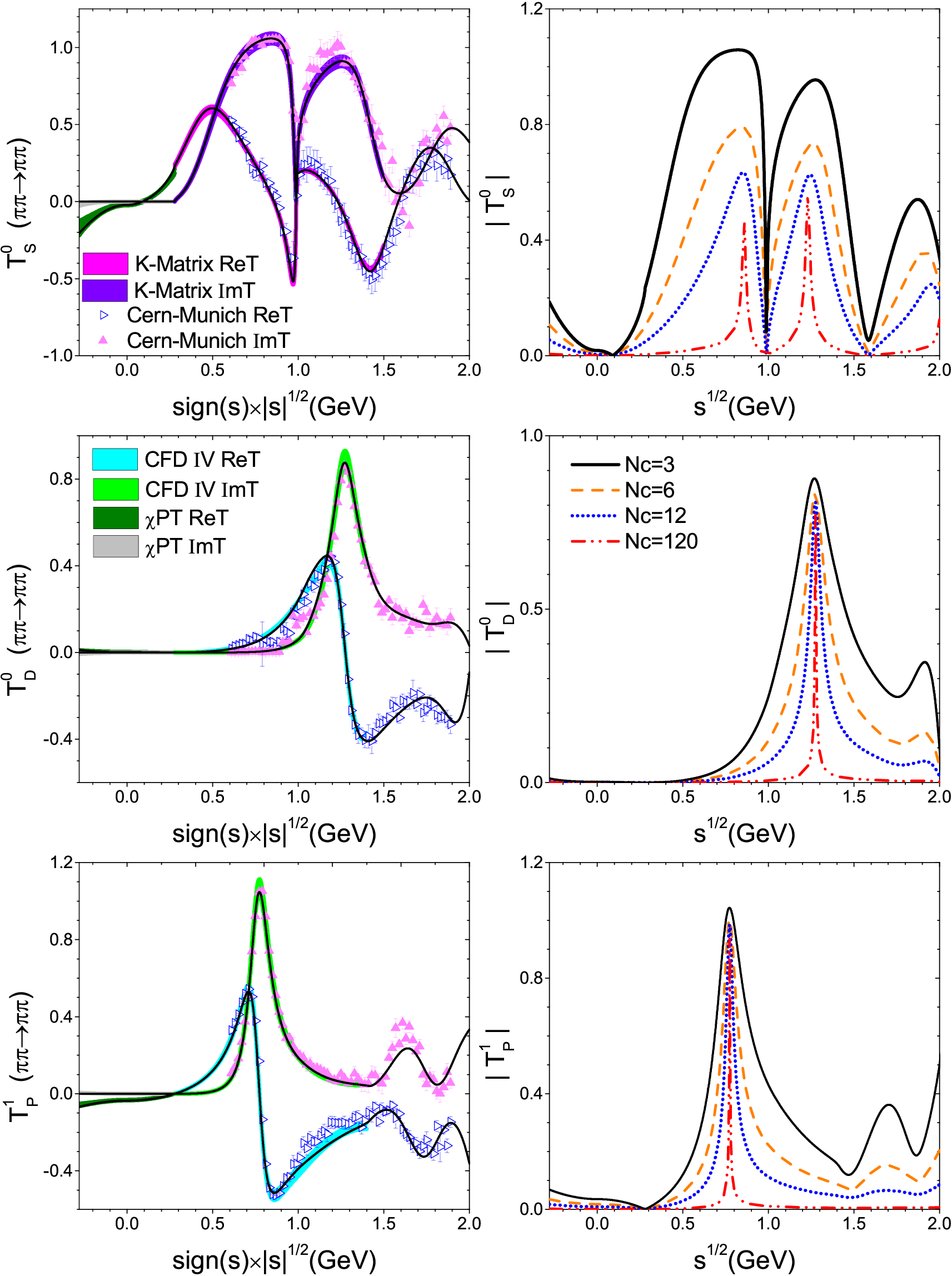}
\caption{\label{Fig:T} The fit of $\pi\pi$ scattering amplitudes are in the left column. The violet and magenta bands are from K-Matrix~\cite{DLY-MRP14}. The olive and light grey bands in the low energy region are from $\chi$PT~\cite{Gasser1984,Gasser1985,Bijnens1994,Pelaez02,DLY11}. The cyan and green bands are from CFDIV~\cite{KPY}.  The CERN-Munich data is from~\cite{CERN-Munich}. The absolute values of the amplitudes by varying $N_C$ are in the right column. The black solid, orange dashed, blue dotted and red dash-dotted lines are for $N_C=3,6,12,120$, respectively. }
\end{figure}

The fit results are shown in Fig.~\ref{Fig:T}. We fit the amplitudes in the region of
$s\in[0,4\rm{GeV}^2]$, where the \lq data' is as follows: $\chi$PT amplitudes at
$[0,4M_\pi^2]$, amplitudes of K-matrix and Roy-like equation at $[4M_\pi^2,2\rm{GeV}^2]$,
and experiment data up to $4\rm{GeV}^2$. The fits are of high quality, even in
our \lq prediction' region where $s\in [-4M_\pi^2,0]$.
In this region the real part of our amplitudes is in good agreement with that
of $\chi PT$ ($\mathcal{O}(p^4)$),
and the imaginary part certainly vanishes. Notice that the imaginary part of the
$\chi PT$ amplitudes is rather small, too.

\begin{figure}[!tbph]
\vspace{-0.0cm}
\includegraphics[width=0.48\textwidth,height=0.6\textheight]{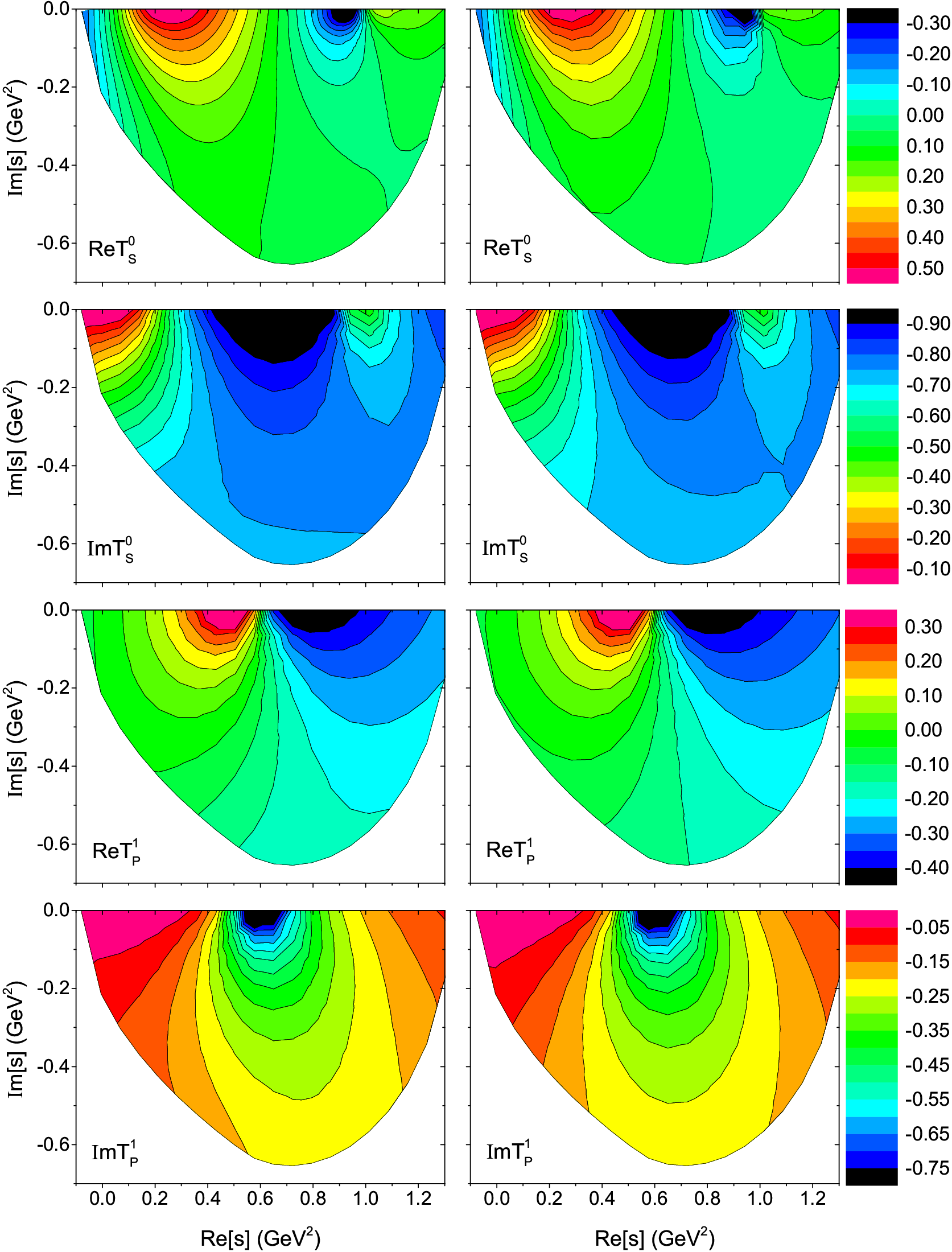}
\caption{Comparison of our amplitudes ($IJ=00,11$) with the ones from the Roy equation
analysis in the
domain where the Roy equations work. On the left side there are real and
imaginary parts of our amplitudes, and on the right side are those from
Roy equations~\cite{colangelo01}.  \label{fig:fit;Roy}}
\end{figure}

There is no l.h.c in our parametrization\footnote{It is worth to note that in
\cite{caprini06} the mass and width of $\sigma$  deviates about $15\%$ from the
original value when the l.h.c is removed.}, but its contribution to the shaded region on
the complex $s$-plane, as shown in Fig.~\ref{fig:fit;Roy}, are properly implemented,
as we fit our amplitudes to the data as well as the amplitudes of Roy equation
\cite{colangelo01} in the presence of crossing symmetry\footnote{The comparison of
the I=1 P-wave amplitudes has already been given in \cite{Dai:2017tew}. However, we
present it here for convenience. Notice that the D-wave is  absent in the  Roy
equation analysis\cite{caprini06}, and the $f_2(1270)$ is quite far away from the l.h.c.
We thus do not discuss it here.}.
We note that the amplitudes on the upper half of $s$-plane are readily obtainable
from the ones on the lower side according to the Schwarz reflection principle.
From Fig.~\ref{fig:fit;Roy} we see that our amplitudes are compatible to that of the
Roy equation analysis in the complex $s$-plane.
The distribution of contours is in good agreement and moreover, their gradient
variations are compatible with each other, as shown by the shading of
the color from blue to red. Nevertheless, amplitudes on the
edge of the domain and in the region of $\rm{Im}s<-0.3$~GeV$^2$ for
$T^0_S(s)$ are less consistent with differences $\leq0.1$.
The difference around $\sqrt{s} \simeq 1$~GeV of  $T^0_S(s)$ are also not ignorable,
this is caused by the different treatment of the $K\overline{K}$: ours
include the physical $K^+K^-$ and $K^0\overline{K}^0$ \cite{DLY-MRP14}, while it
is treated as  $K\overline{K}$ in the isospin basis in the analysis of the Roy
equations \cite{colangelo01}.

With these amplitudes, we can extract the poles and residues on the second sheet.
The residue $g_{f\pi\pi}$ and pole $s_R$ on the second Riemann sheet are defined as
\bea
T^{II}(s)=\frac{g_{f\pi\pi}^{2}}{s_R-s}\,.
\label{eq;g}
\eea
The pole locations and their residues are listed in Table~\ref{tab:poles}.
\begin{table}[tbp]
{\footnotesize
\begin{tabular}{|c || c |c |@{}c @{}|@{}c@{} |}
\hline
\rule[-0.3cm]{0cm}{0.8cm}\multirow{2}{*}{\rule[-1cm]{0cm}{2cm}State}  &  pole locations  &\multicolumn{2}{c|}{$g_{f\pi\pi}=|g_{f\pi\pi}|e^{i\varphi}$~}  \\
\cline{3-4}
\rule[-0.3cm]{0cm}{0.8cm}    & (MeV)   & $~|g_{f\pi\pi}|~({\rm GeV})~$ & $~\varphi~(^\circ)~$   \\
\hline\hline
$\sigma/f_0(500)$ & $436.2(12.2)-i260.7(6.8)$
     & $0.45(0.02)$   & $74(2)$      \rbox \\[0.5mm] \hline
$f_0(980)$        & $997.7(1.1)-i21.7(1.9)$
     & $0.27(0.02)$   & $84(3)$    \rbox \\[0.5mm] \hline
$f_0(1370)$        & $1431.6(34.6) -i185.4(22.4)$
     & $0.78(0.21)$   & $-47(18)$    \rbox \\[0.5mm] \hline
$f_2(1270)$       & $1278.3(7.0) -i79.3(17.8)$
        & $0.50(0.05)$   &~$0.7(4.3)$    \rbox \\[0.5mm] \hline
$\rho(770)$       & $762.4(3.9) -i68.7(6.3) $
    & $0.34(0.01)$   & $12(3)$                     \rbox \\[0.5mm] \hline
\end{tabular}
\caption{\label{tab:poles}The pole locations and residues on the second Riemann sheet.  }
}
\end{table}
These are very similar from those of previous
analyses~\cite{caprini06,PelaezPRL,DLY-MRP14,Rusetsky2011,PDG16}.
For the $f_0(1370)$ and $f_2(1270)$, to find the pole closest to the physical sheet
one needs to include the $\pi\pi$, $\overline{K}K$, 4$\pi$ as coupled channels.
The pole obtained here from a single $\pi\pi$ channel is of course,
not on the Riemann sheet which is closest to the physical region. Therefore, we do
not discuss the $f_0(1370)$ in the next sections.
For the $f_2(1270)$, $\pi\pi$ is the dominant decay channel and the pole
on the second sheet is not far away from the physical one.

Having analytically calculated the partial wave amplitudes of $IJ=00,11,02$ within
one-loop SU(3) $\chi$PT, we can  match our dispersive results to these and so fix
their $N_C$ dependence.
We note several points about this matching:
First, though the matching is done in the low-energy region,
this $N_C$ dependence is transferred to  high-energy region.
As the FSI of hadrons in the higher energy region correspond
to  \lq hadron loop' corrections, which could be translated into  higher
order corrections (quark loops) in large $N_C$ QCD,
and are thus suppressed by an extra factor of $N_C^{-1}$. Of course, this $N_C$
dependence is energy-dependent, see the discussion below.
Second, we only calculate the $\chi$PT amplitudes up to $\mathcal{O}(p^4)$.
Here, the imaginary part of the $\chi$PT amplitudes is given entirely from
one-loop integrals and thus both l.h.c and r.h.c are $N_C^{-2}$,
while the real part is $N_C^{-1}$.
For $\mathcal{O}(p^6)$, it is still $N_C^{-1}$ for the
real part of the amplitudes as given by the contact terms, and again it is $N_C^{-2}$ for the imaginary part.
The latter one is either from the 2-loop corrections or from the 1-loop corrections with one insertion of
a  contact term of $\mathcal{O}(p^4)$.
This $N_C$ dependence continues when one goes to higher orders. One thus finds
that $\rm{ImT}/\rm{ReT}\sim 1/N_C$ and we define:
\be
\varphi(s,N_C) \;=\; \arctan \left[\frac{3}{N_C}\tan \varphi(s) \right] \;, \label{eq:phi;Nc}
\ee
where $\varphi(s)$ is the phase. One notices that $\varphi(M_R^2,N_C)=90^\circ$, where
the phase is $90^\circ$ at $M_R$ with $N_C=3$.
This implies for the the low-energy region:
\bea\label{eq:Omnes;expand}
\Omega^I_J(s,N_C)=\bar{\Omega}^I_J(s)+\mathcal{O}(N_C^{-1})\;.
\eea
$\bar{\Omega}^I_J(s)$ is close to 1, where the phase is obtained by
$N_C\to\infty$ and behaves as a step function.
Third, we only match up to $\mathcal{O}(N_C^{-1})$. In this case,
Eq.(\ref{eq:Omnes;expand}) shows that one only needs to inject
the $N_C$ dependence as
\be
P^I_J(s,N_C)\;=\;\frac{3}{N_C}P^I_J(s)\;. \label{eq:P;Nc}
\ee
This is compatible with the fact that the real part of the $\chi$PT amplitudes
is $\mathcal{O}(N_C^{-1})$ while the imaginary part of the $\chi$PT amplitudes
is $\mathcal{O}(N_C^{-2})$, up to any order. For simplicity we ignore all the
complicated higher order $N_C$ dependence.

\begin{figure}[ht]
\includegraphics[width=0.48\textwidth,height=0.6\textheight]{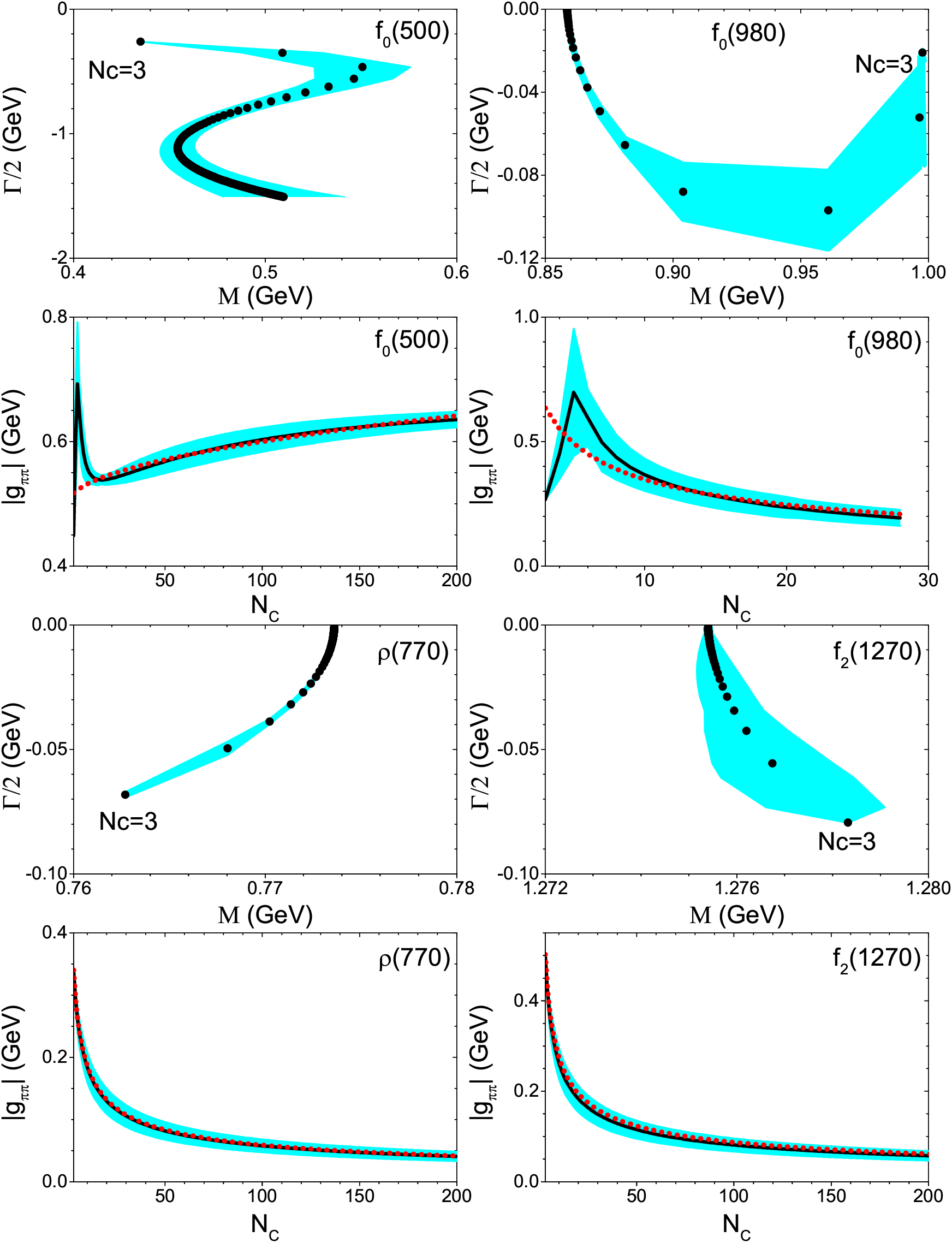}
\caption{\label{Fig:pole} The trajectories of the pole locations and their residues by
varying $N_C$ from 3 to a large number, in steps of $\Delta N_C=1$.
The black filled circle and lines represent poles and residues, respectively.
The red dotted lines are linear fits (with the variable $\sqrt{N_C}$ or 1/$\sqrt{N_C}$)
to the residues obtained, as explained in the text.}
\end{figure}

We test the stability of the amplitudes by varying $N_C$, shown in Fig.~\ref{Fig:T}.
Both the shape and the magnitude of the amplitude of the $S$ wave vary dramatically.
The peak of $|T|$ behaves as $\mathcal{O}(1)$ for the $P$ and $D$ waves and the
width behaves as $\mathcal{O}(N_C^{-1})$. This is consistent with that of a
Breit-Wigner description, where there is a zero of the real part amplitude ($s=M_R^2$),
and the $N_C$ dependence is canceled leading to
$T\simeq i/\rho$. Thus, the $N_C$ dependence is
properly implemented.
The trajectories of the pole locations and their residues $g_{f\pi\pi}$ on
the second Riemann sheet are obtained from Eq.~(\ref{eq;g}) and plotted in Fig.~\ref{Fig:pole}.
To estimate the uncertainties, we choose $\frac{2}{N_C}+\frac{3}{N_C^2}$ and/or
$\frac{4}{N_C}-\frac{3}{N_C^2}$ to replace $\frac{3}{N_C}$ in
Eqs.~(\ref{eq:phi;Nc},\ref{eq:P;Nc}). The variation of poles and residues with
these different $N_C$ dependence are collected as the uncertainty.
Here we simply assume that the whole contribution of the $N_C^{-2}$ corrections is
roughly one third of that of $N_C^{-1}$.
The separated $N_C^{-2}$ dependence of each polynomial in Eq.~(\ref{eq:P;Nc}) is
not discussed, as the complete $N_C^{-2}$ part of the real amplitudes in
$\chi PT$ is not available and these polynomials are correlated with each other.
Only the first two polynomials are undoubtedly fixed due to the scattering lengths
and slope parameters. We are aware that the $N_C$ dependence of these polynomials,
especially the third or fourth polynomial could have fairly large effects on
the uncertainties. One needs a careful analysis on $N_C^{-2}$ corrections
for a refined analysis.
The uncertainties of the poles/residues are presented as cyan bands in Fig.~\ref{Fig:pole}.
They are compatible to the black solid line with only  $N_C^{-1}$ dependence\footnote{We
note that Ref.~\cite{Jacob2015} points out the sub-leading $N_C$ dependence of the
LECs may be sizable. In \cite{Pelaez04,Nieves:2009kh}, the uncertainty caused
by the regularization scale $\mu$ is discussed. These uncertainties are larger
than that shown in Fig.\ref{Fig:pole}, particularly for scalar mesons. Here, we
simply absorb all these uncertainties into an overall higher correction of
$\mathcal{O}(N_C^{-2})$.  }.
Next, we discuss the various resonances within the accuracy of our approach.

\vspace{1mm}
\noindent $\mathbf{\rho(770)}$\\
The pole trajectory of the $\rho(770)$ moves toward the real axis, similar to what was
found using U$\chi$PT~\cite{Pelaez06,DLY11,MRP11}.
In Fig.\ref{Fig:pole}, the red dotted line represents for
$|g_{\rho\pi\pi}|\times\sqrt{3/N_C}$ with $|g_{\rho\pi\pi}|$ the residue at $N_C=3$. The black solid and red dotted lines overlap
perfectly. This confirms that the modulus of the residue decreases exactly as $1/\sqrt{N_C}$. Similarly the width decreases exactly as $1/N_C$.
Such behavior confirms the widely accepted $\bar{q}q$ structure.

\vspace{1mm}
\noindent $\mathbf{f_2(1270)}$\\
For the $f_2(1270)$, the trajectory is quite similar to that of the $\rho(770)$.
Again this confirms a $\bar{q}q$ structure.
In Fig.\ref{Fig:pole}, the red dotted line represents
$|g_{f_2\pi\pi}|\times\sqrt{3/N_C}$ with $|g_{f_2\pi\pi}|$ the residue at $N_C=3$. The black solid and red dotted lines are much the same,
indicating that the residue decreases exactly as $1/\sqrt{N_C}$. Similarly, the width decreases exactly as $1/N_C$.

\vspace{1mm}
\noindent $\mathbf{\sigma/f_0(500)}$\\
For $\sigma/f_0(500)$, its mass is roughly $\mathcal{O}(1)$ and the width is
$\mathcal{O}(N_C)$.
Mixing with $\mathcal{O}(N_C)$ or  $\mathcal{O}(N_C^{-1})$ can not be excluded.
The pole moves far away from the real axis on the complex s-plane when $N_C$
is increased. We note that the third and fourth quadrants in Eq.~(\ref{eq:P}) may
be important and their higher $N_C^{-2}$ corrections may distort this trajectory.
However, considering that the pole is mainly determined by the unitary cut at
$N_C=3$, and both the l.h.c and distant r.h.c are $\mathcal{O}(N_C^{-2})$, and
also in the $\mathcal{O}(N_C^{-1})$ case the pole moves far away from the real axis
(or cuts), it is natural to assume that the $1/N_C$ part of these quadrants dominates the
trajectory at large $N_C$. At present, we can not quantify how much difference
these $\mathcal{O}(N_C^{-2})$ corrections would generate.
To get such a large width it could have a molecular component~\cite{Knecht2013} and/or
a tetra-quark component \cite{Cohen:2014vta}.
Notice in~\cite{DLY11,DLY-MRP14} the shadow pole in the third sheet
suggests a $\bar{q}q$ component.
So the $\sigma$ might be a mixed state including molecule,
$\bar{q}q$, etc. The modulus of the residue increases, reaching the peak at
$N_C=5$ rapidly and then
decreases, again it increases much slower at the turning point $N_C=17$.
The red dotted line is for $0.47+0.013\sqrt{N_C}$, and it is essentially coincident
with the black solid line where $N_C\geq17$.
This indicates that the residue should contain components with
$\mathcal{O}(N_C^{1/2})$ and $\mathcal{O}(1)$.
Also, the sharp peak implies that the residue should contain $\mathcal{O}(N_C)$ or higher components. 
A mixture of $\mathcal{O}(N_C^{-1/2})$ or lower components is also possible.
Moreover, it is of interest to note that around the sharp peak (with $N_C\in[3,7]$)
the real parts of the pole locations in complex s-plane are quite close to the
$\pi\pi$ threshold. The curved behavior of the trajectory is consistent with
the mixing structure of molecule and $\bar{q}q$, while an extra tetra-quark
component is possible.

\vspace{1mm}
\noindent $\mathbf{f_0(980)}$\\
For the $f_0(980)$, the pole moves to the real axis, slightly below $\bar{K}K$
threshold. It is similar to that of $\rho(770)$ and $f_2(1270)$, implying an
$\bar{s}s$ component. In contrast, in Refs~\cite{Uehara04,DLY11} the pole moves to
the real axis above $\bar{K}K$ threshold and goes onto the fourth Riemann sheet.
We may need higher order $1/N_C$ corrections, especially that caused by kaon loops,
to obtain a more accurate pole trajectory.
The residue increases at first and then decreases, as a \lq curve'. The red dotted
line is $1.10/\sqrt{N_C}$, which is coincident with the black solid line where $N_C\geq10$.
This indicates that the residue must have the component of $\mathcal{O}(N_C^{-1/2})$.
Note it is most likely to be a $\overline{K}K$ molecule in other
analysis~\cite{MRP10,DLY11,DLY-MRP14}.
The peak around $N_C=5$ implies also $\mathcal{O}(N_C^{1/2})$ or even higher components.
This might indicate a possible tetra-quark component.
We note that the pole locations, around the sharp peak with $N_C\in[3,6]$, are close
to the $\overline{K}K$ threshold.
Our findings support the idea that the $f_0(980)$ is a mixture of $\overline{K}K$ molecular and
$\bar{s}s$ components. The relative strengths of these components can not be inferred
from the analysis presented here.

In this letter we presented a new method to study the large $N_C$ behaviour of resonances.
The $\pi\pi$ scattering amplitudes with FSI are constructed in a model-independent way.
We match it with the amplitudes of $\chi$PT to give the
$N_C$-dependence of the coefficients and phase in the dispersive approach. This is a
reliable way to study the $N_C$-dependence with final-state interactions.
We obtain the trajectories of the poles and residues as $N_c$ changes.
Those for both the $\rho(770)$ and the
$f_2(1270)$ quantitatively support, as expected, the standard
$\bar{q}q$ structure.
The $N_C$ trajectory of the light scalar mesons are obtained,
especially the coupling of $g_{\sigma\pi\pi}$ has $\mathcal{O}(N_C^{1/2})$ component
and $g_{f_0(980)\pi\pi}$ has $\mathcal{O}(N_C^{-1/2})$ component.
They are consistent with each being a combination of $\bar{q}q$ and multi-hadron
(molecular) components, while a tetra-quark component can not be excluded.
We stress that some of these conclusions might be modified when higher order
corrections in the $\chi$PT amplitudes are accounted for.

\section*{Acknowledgements}
We are very grateful to M.R. Pennington for many valuable suggestions and discussions
to improve the paper. Special thanks to J. Ruiz de Elvira for his thoughtful and
critical reading of the manuscript. Helpful discussions with Z.H. Guo  are also acknowledged.
This work is supported by the DFG (SFB/TR 110, ``Symmetries and the Emergence of
Structure in QCD''), by the Chinese Academy of Sciences (CAS) President's
International Fellowship Initiative (PIFI) (Grant No. 2018DM0034) and the VolkswagenStiftung
(Grant No. 93562).

\end{document}